\documentclass{jdfsl}

\usepackage{stfloats} 

\let\cite\shortcite
\let\citeA\shortciteA

\newcommand\blfootnote[1]{%
  \begingroup
  \renewcommand\thefootnote{}\footnote{#1}%
  \addtocounter{footnote}{-1}%
  \endgroup
}
\usepackage{graphicx}

\usepackage{url}
\usepackage{balance}
\usepackage{tabularx}
\newcolumntype{L}[1]{>{\raggedright\arraybackslash}p{#1}}
\newcolumntype{C}[1]{>{\centering\arraybackslash}p{#1}}
\newcolumntype{R}[1]{>{\raggedleft\arraybackslash}p{#1}}
\newcommand{\Tstrut}{\rule{0pt}{2.6ex}}         
\newcommand{\tline}{\hline\Tstrut}
\newcommand{\mline}{\hline\Tstrut}
\newcommand{\bline}{\hline}

\begin{document}

\title{Hierarchical Bloom Filter Trees for Approximate Matching}


\author{
	David Lillis \\ Forensics and Security Research Group \\ School of Computer Science \\ University College Dublin \\\url{david.lillis@ucd.ie} \and
	Frank Breitinger \\Cyber Forensics Research \\ and Education Group \\ Tagliatela College of Engineering \\ University of New Haven \\ \url{fbreitinger@newhaven.edu} \and
	Mark Scanlon\\ Forensics and Security Research Group \\ School of Computer Science \\ University College Dublin \\\url{mark.scanlon@ucd.ie} }

\institute{}

\abstract{
Bytewise approximate matching algorithms have in recent years shown significant promise in detecting files that are similar at the byte level. This is very useful for digital forensic investigators, who are regularly faced with the problem of searching through a seized device for pertinent data. A common scenario is where an investigator is in possession of a collection of ``known-illegal'' files (e.g. a collection of child abuse material) and wishes to find whether copies of these are stored on the seized device. Approximate matching addresses shortcomings in traditional hashing, which can only find identical files, by also being able to deal with cases  of merged files, embedded files, partial files, or if a file has been changed in any way.

Most approximate matching algorithms work by comparing pairs of files, which is not a scalable approach when faced with large corpora. This paper demonstrates the effectiveness of using a ``Hierarchical Bloom Filter Tree'' (HBFT) data structure to reduce the running time of collection-against-collection matching, with a specific focus on the \texttt{MRSH-v2} algorithm. Three experiments are discussed, which explore the effects of different configurations of HBFTs. The proposed approach dramatically reduces the number of pairwise comparisons required, and demonstrates substantial speed gains, while maintaining effectiveness.
}


\maketitle

\section{Introduction}\label{sec:introduction}

Current digital forensic process models are surprisingly arduous, inefficient, and expensive.\blfootnote{This paper is an extended version of \citeA{Lillis2017}, which was presented at the 9th EAI International Conference on Digital Forensics and Cyber Crime (ICDF2C), Prague, Czech Republic, 9-11 October, 2017.} Coupled with the sheer volume of digital forensic investigations facing law enforcement agencies worldwide, this has resulted in significant evidence backlogs becoming commonplace~\cite{scanlon2016deduplication}, frequently reaching 18-24 months~\cite{casey2009investigation} and exceeding 4 years in extreme cases~\cite{lillis2016challenges}. The backlogs have grown due to a number of factors including the volume of cases requiring analysis, the number of devices per case, the volume of data on each device, and the limited availability of skilled experts~\cite{quick2014impacts}. Automated techniques are in continuous development to aid investigators, but due to the sensitive nature of this work, the ultimate inferences and decisions will always be made by skilled human experts~\cite{james2015automated}.

Perhaps the most common (and most time-consuming) task facing digital investigators involves examination of seized suspect devices to determine if pertinent evidence is contained therein. Often, this examination requires significant manual, expert data processing and analysis during the acquisition and analysis phases of an investigation. A number of techniques have been created or are in development to expedite/automate parts of the typical digital forensic process. These include triage~\cite{rogers2006computer}, distributed processing~\cite{roussev2004breaking}, Digital Forensics as a Service (DFaaS)~\cite{vanBaar2014S54}, workflow management and automation~\cite{braekt, gupta2016heuristic}. While these techniques can help to alleviate the backlog, the premise behind many of them involves evidence discovery based on exact matching of hash values (e.g., MD5, SHA1). Typically, this requires a set of hashes of known incriminating/pertinent content. The hash of each artefact from a suspect device is then compared against this set. This approach falls short against basic counter-forensic techniques (e.g., content editing, content embedding, data transformation).

Approximate matching (also referred to as ``fuzzy hashing'') is one technique used to aid the discovery of these obfuscated files~\cite{breitinger2014approximate}. A number of algorithms have been developed including \texttt{ssdeep}~\cite{Kornblum2006}, \texttt{sdhash}~\cite{Roussev2010}, and \texttt{MRSH-v2}~\cite{Breitinger2012}. This paper focuses specifically on \texttt{MRSH-v2}. This algorithm operates by generating a ``similarity digest'' for each file, represented as Bloom filters \cite{bloom1970space}. An all-against-all pairwise comparison is then required to determine if files from a set of desired content is present in a corpus of unanalysed digital material. Thus, \texttt{MRSH-v2} does not exhibit strong scalability for use with larger datasets.

This paper presents an improvement in the runtime efficiency of approximate matching techniques, primarily through the implementation of a Hierarchical Bloom Filter Tree (HBFT). Additionally, it examines some of the tunable parameters of the algorithm to gauge their effect on the required running time. A number of experiments were conducted, using two different formulations of a HBFT, which indicated a substantial reduction in the running time, in addition to which the final experiment achieved a 100\% recall rate for identical files and also for files that have a \texttt{MRSH-v2} similarity above a reasonable threshold of 40\%.

Section~\ref{sec:background} outlines the prior work that has been conducted in the area of approximate matching. The operation of \texttt{MRSH-v2} is discussed in Section~\ref{sec:mrsh-v2}. HBFTs are introduced in Section~\ref{sec:hbft}. Section~\ref{sec:experiments} presents the series of experiments designed to evaluate the effectiveness of the HBFT approach, and finally Section~\ref{sec:conclusions} concludes the paper and outlines directions for further work.

\section{Background: Approximate Matching} \label{sec:background}

Bytewise approximate matching for digital forensics gained popularity in 2006 when \citeA{Kornblum2006} presented context-triggered piecewise hashing (CTPH) including an implementation called \texttt{ssdeep}. It was at that time referred to as ``fuzzy hashing''. Later, this term converted to ``similarity hashing'' (most likely due to \texttt{sdhash} which stands for ``similarity digest hash''~\cite{Roussev2010}). In 2014, the National Institute of Standards and Technology (NIST) developed Special Publication 800-168, which outlines the definition and technology for these kinds of algorithms~\cite{breitinger2014approximate}.

In addition to the prominent aforementioned implementations, there are several others. \texttt{MinHash}~\cite{broder1997resemblance} and \texttt{SimHash}~\cite{sadowski2007simhash} are ideas on how to detect/identify small changes (up to several bytes), but were not designed to compare hard disk images with each other. \citeA{oliver2013tlsh} presented an algorithm named \texttt{TLSH}, which is premised on locality sensitivity hashing (LSH). There are significantly more algorithms, but to explain all of them would be beyond the scope of this paper; a good summary is provided by \citeA{harichandran2016bytewise}.

While these algorithms have great capabilities, they suffer one significant drawback, which we call the ``database lookup problem''. In comparison to traditional hash values which can be sorted and have a lookup complexity of $O(1)$ (hashmap) or $O(log(n))$ (binary tree; where $n$ is the number of entries in the database), looking up a similarity digest usually requires an all-against-all comparison ($O(n^2)$) to identify all matches. To overcome this drawback, \citeA{Breitinger2014b} presented a new idea that overcomes the lookup complexity (it is approximately $O(1)$) but at the cost of inaccuracy. More specifically, the method allows item vs. set queries, resulting in the answer either being ``yes, the queried item is in the set'' or ``no, it is not''; one cannot say against which item it matches.

As a means of addressing these drawbacks, \citeA{Breitinger2014} presented a further article where they offered a theoretical solution to the lookup problem, based on a tree of Bloom filters. However, an implementation (and thus a validation) has not been conducted to date. We refer to this as a Hierarchical Bloom Filter Tree (HBFT). The focus of the present work is the empirical evaluation of this approach, so as to demonstrate its effectiveness and to investigate some practical factors that affect its performance.

\section{The \texttt{MRSH-v2} algorithm} \label{sec:mrsh-v2}

The work in this paper is intended to improve upon the performance of the \texttt{MRSH-v2} algorithm. Therefore, it is important to firstly outline its operation in informal terms, which will aid the discussion later. A more detailed, formal description of the algorithm can be found in the paper by \citeA{Breitinger2012}. The primary goal of \texttt{MRSH-v2} is to compress any byte sequence and output a similarity digest. Similarity digests are created in a way that they can be compared with each other, which will result in a similarity score. Each similarity digest is a collection of Bloom filters~\cite{bloom1970space}.

To create the similarity digest, \texttt{MRSH-v2} splits an input into chunks (also known as ``subhashes'') of approximately 160 bytes. These chunks are hashed using \texttt{FNV} (a fast non-cryptographic hash function), which is used to set 5 bits of the Bloom filter. To divide the input into chunks, it uses a window of 7 bytes, which slides through the input byte-by-byte. The content of the window is processed and whenever it hits a certain value (based on a modulus operation), the end of a chunk is identified. Thus, the actual size of each chunk varies. Each Bloom filter has a specific capacity. Once this has been reached, any further chunks are inserted into a new Bloom filter that is appended to the digest. Approximate matching occurs by comparing similarity digests against one another. To compare two file sets, an all-against-all pairwise comparison is required.

One way to improve upon the all-against-all comparison is to use the file-against-set strategy outlined by \citeA{Breitinger2014b}. An alternative strategy that has not yet been fully evaluated is to use a hierarchical Bloom filter tree (HBFT), as suggested by \citeA{Breitinger2014}. This approach is intended to achieve speed benefits over a pairwise comparison while supporting the identification of specific matching files. The primary contribution of this paper is to investigate the factors that affect the runtime performance of this latter approach, compared to the pairwise comparisons required by the original algorithm.

\section{Hierarchical Bloom Filter Trees (HBFT)} \label{sec:hbft}

\begin{figure*}[tb]
\centering
\includegraphics[width=\textwidth]{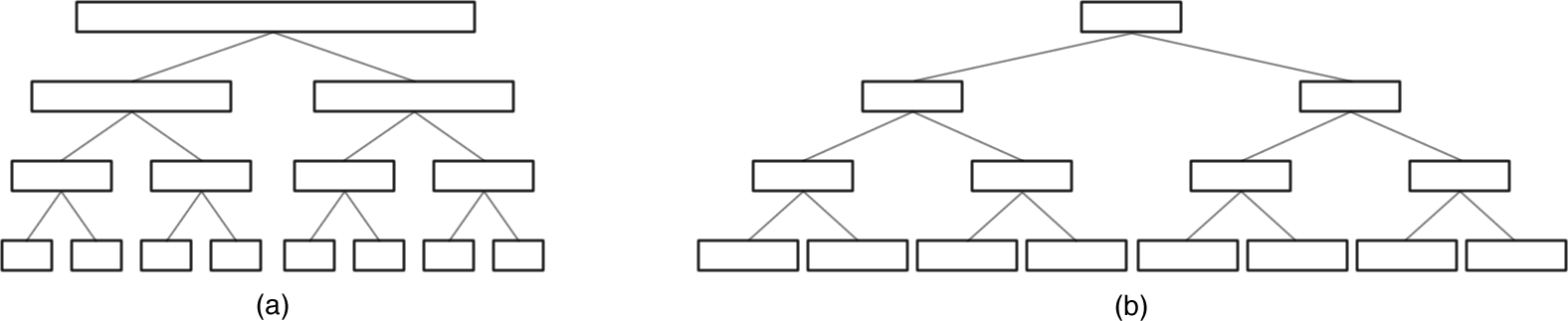}
\caption{Hierarchical Bloom Filter Tree (HBFT) structure using (a) variable-width and (b) fixed-width Bloom filters.}
\label{fig:treeoverview}
\end{figure*}

In a Hierarchical Bloom Filter Tree (HBFT), the root node of the tree is a Bloom filter that represents the entire collection. A key feature of a Bloom filter is that it can say only whether an item is probably contained in it, or definitely not contained in it. Thus it is possible to give false positive results, but not false negatives. The rate of false positives depends on the size of the Bloom filter and the number of items it contains. When searching for a file, if a match is found at the root of the tree, its child nodes can then be searched. Although this structure is inspired by a classic binary search tree, a match at a particular node in a HBFT does not indicate whether the search should continue in the left or right subtree. Instead, both child nodes need to be searched, with the search path ending when a leaf node is reached or a node does not match.

Two forms of tree layout are shown in Figure~\ref{fig:treeoverview}: one uses Bloom filters of different sizes (referred to as a ``variable-width'' HBFT), whereas the other uses a single fixed size for each Bloom filter. For the variable-width tree, each level in the tree is allocated an equal amount of memory. Thus each Bloom filter occupies half the memory of its parent, and also represents a file set that is half the size of its parent. The expected false positive rates will be approximately equal at all levels in the tree. In contrast, a fixed-width tree uses the same size for every Bloom filter. Thus each level of the tree occupies twice the space of the level above.

When a collection is being modelled as a HBFT, each file is inserted into the Bloom filter at some leaf node in the tree, and also into its ancestor nodes. The mechanism of inserting a file into a Bloom filter is the same as for the single Bloom filter approach outlined by \citeA{Breitinger2014b}, which is also very similar to the approach taken by the classic \texttt{MRSH-v2} algorithm outlined in Section~\ref{sec:mrsh-v2}. The key difference is that instead of creating a similarity digest of potentially multiple small Bloom filters for an individual file, each subhash is used to set 5 bits of the larger Bloom filter within a tree node that usually relates to multiple files. 

Depending on the design of the tree, a leaf node may represent multiple files. Thus a search that reaches a leaf node will still require a pairwise comparison with each file in this subset, using \texttt{MRSH-v2}. However, given that most searches will reach only a subset of the root nodes, the number of pairwise comparisons required for each file is greatly reduced.

The process to check if a file matches a Bloom filter node is similar to the process of inserting a file into the tree. However, instead of inserting each hash into the node, its subhashes are instead checked against the Bloom filter to see if they are contained in it. If a specific number of consecutive hashes are contained in the node, this is considered to be a match. The number of consecutive hashes is configurable as a parameter named \texttt{min\_run}. The first experiment in this paper (discussed in Section~\ref{sec:experiment1}) explores the effects of altering this value.

In constructing a HBFT, memory constraints will have a strong influence on the design of the tree. In practical situations, a typical workstation is unlikely (at present) to have access to over 16GiB of main memory. Thus trade-offs in the design of the tree are likely. Larger Bloom filters have lower false positive rates (assuming the quantity of data is constant), but lead to shallower trees (thus potentially increasing the number of pairwise comparisons required).

Of the two proposed designs for a HBFT, each has its own theoretical advantages. One aim of the ensuing experiments is to identify if any of these has more influence in practice. Some considerations worthy of note include:

\begin{itemize}
   \item Calculating the union of two Bloom filters of equal size is trivially performed using a bitwise-OR operation. Thus a fixed-width HBFT can be constructed in a bottom-up manner, whereby each file needs only be inserted into a leaf Bloom filter. Once all files have been processed, these leaves can be recursively merged to create the parent nodes. In contrast, a variable-width tree design requires each file to be inserted separately into an appropriate node at every level in the tree.
   \item Another consequence of the above observation is that if the HBFT is to be distributed over multiple computational nodes, this has consequences for the quantity of data that must be shared between nodes when building the tree. Instead of sending the hashes of all files throughout the system, Bloom filters can be shared and locally merged where necessary. This is outside of the scope of this paper, but is discussed as future work in Section~\ref{sec:conclusions}.
   \item The memory required at each level of a fixed-width tree increases exponentially. This means that for an equal amount of available memory, a fixed-width tree must necessarily either be built to be shallower than the variable-width tree, or make use of smaller Bloom filters towards the top of the tree. The latter approach results in a higher false-positive rate in this part of the tree, which will likely lead to deeper searches for files that are not matched with anything in the corpus. If the size of the leaf nodes is equal, then the overall false positive rate of the two trees will be equivalent.
\end{itemize}

For both types of tree, larger Bloom filters result in lower false positive rates at the expense of a shallower tree (since memory is limited). In a shallower tree, each leaf node represents a larger subset of the corpus, which may require a greater number of pairwise comparisons for each search.
 
\section{Experiments} \label{sec:experiments}

As part of this work, a number of experiments were conducted to examine the factors that affect the performance of the HBFT structure. In each case, a HBFT was used to model the contents of a dataset. Files from another dataset were then searched for in the tree, and the results reported. Because the speed of execution is of paramount importance, and because the original \texttt{MRSH-v2} implementation was written in C, the HBFT implementation used for these experiments was also written in that language. The source code has been made available under the Apache 2.0 licence\footnote{Available at \url{http://github.com/ishnid/mrsh-hbft}}.

The workstation used for the experiments contains a quad-core Intel Core i7 2.67GHz processor, 12GiB of RAM and uses a solid state drive for storage. The operating system is Ubuntu Linux 16.04 LTS. The primary constraint this system imposes on the design of experiments is that of the memory that is available for storing the HBFTs. For all experiments, the maximum amount of memory made available for the HBFT was 10GiB. The size of the individual Bloom filters within the trees then depended on the number of nodes in the tree (which in turn depends on the number of leaf nodes).

For each experiment, the number of leaf nodes ($n$) is specified in advance, from which the total number of nodes can be computed (since this is a complete binary tree). Given the upper total memory limit ($u$, in bytes), and that the size of each Bloom filter (in bytes) should be a power of two (per~\citeA{Breitinger2014b}), it is possible to calculate the maximum possible size of each Bloom filter.

For a variable-width tree, all levels in the tree are allocated the same amount of memory. Therefore the size of the root Bloom filter in bytes ($r$) is given by:
\begin{equation}
r = 2^{\lfloor log_2(u/(log_2(n)+1))\rfloor}
\end{equation}

The size of the other nodes in bytes is then $\frac{r}{2^d}$ where $d$ is the depth of the node in the tree (i.e. the size of a Bloom filter is half the size of its parent).

For a HBFT with fixed-sized Bloom filters, all nodes have size equal to that of the root node. Here, $r$ is given by:
\begin{equation}
r = 2^{\lfloor log_2(u/(2n - 1))\rfloor}
\end{equation}

The ultimate goal of the experiments is to demonstrate that the HBFT approach can improve the running time of an investigation over the all-against-all comparison approach of \texttt{MRSH-v2} without suffering a degradation in effectiveness. It achieves this by narrowing the search space so that each file that is searched for need only be compared against a subset of the dataset.

Using a HBFT, the final outcome will be a set of similarity scores. This score is calculated by using \texttt{MRSH-v2} to compare the search file with all files contained in any leaves that are reached during the search. Therefore, the HBFT approach will not identify a file as being similar if \texttt{MRSH-v2} does not also do so.

In these experiments, the similarity scores generated by \texttt{MRSH-v2} are considered to be ground truth. Evaluating the degree to which this agrees with the opinion of a human judge, or how it compares with other  algorithms, is outside the scope of this paper. The primary difference between the outputs is that the HBFT may fail to identify files that \texttt{MRSH-v2} considers to be similar (i.e. false negatives) due to an appropriate leaf node not being reached. 

Therefore the primary metric used, aside from running time, is recall: the proportion of known-similar (or known-identical) files for which the HBFT search reaches the appropriate leaf node.

\subsection{Datasets}

Two datasets were used as the basis for the experiments conducted in this paper:
\begin{itemize}
	\item The \textbf{t5} dataset~\cite{Roussev2011} is frequently used for approximate matching experimentation. It consists of 4,457 files (approximately 1.8 GiB) taken from US government websites. It includes plain text files, HTML pages, PDFs, Microsoft Office documents and image files.
	\item The \textbf{win7} dataset is a fresh installation of a Windows 7 operating system, with default options selected during installation. It consists of 48,384 files (excluding symbolic links and zero-byte files) and occupies approximately 10GiB.
\end{itemize}

The first two experiments use one or both of these datasets directly. The final experiment includes some modifications, as outlined in Section~\ref{sec:experiment3}.

\subsection{Experiment Overview}

The following sections present three experiments that were conducted to evaluate the HBFT approach. Section~\ref{sec:experiment1} compares the \texttt{t5} dataset with itself. This is intended to find whether the HBFT approach is effective in finding identical files, and to investigate the effect of varying certain parameters when designing and searching a HBFT. It also aims to demonstrate the extent to which the number of pairwise comparisons required can be reduced by using this technique.

Section~\ref{sec:experiments2} uses disjoint corpora of different sizes (\texttt{t5} and \texttt{win7}). In a typical investigation, there may be a large difference between the size of the collection of search files and a seized hard disk. This experiment aims to investigate whether it is preferable to use the tree to model the smaller or the larger corpus. Additionally, it examines the performance characteristics of fixed and variable width HBFTs.

Finally, Section~\ref{sec:experiment3} uses overlapping corpora where a number of files have been planted on the disk image. These files are identical to, or similar to, files in the search corpus. This experiment demonstrates that using a HBFT is substantially faster than the pairwise approach.

\subsubsection{Experiment 1: \texttt{t5} vs. \texttt{t5}} \label{sec:experiment1}

For the initial experiment, the HBFT was constructed to represent the \texttt{t5} corpus. All files from \texttt{t5} were also used for searching. Thus every file searched for is also located in the tree and should be found. Conducting an all-against-all pairwise comparison using \texttt{MRSH-v2} required a total of  19,864,849 comparisons, which took 319 seconds.

\begin{figure*}[b]
	\centering
	\includegraphics[width=225pt]{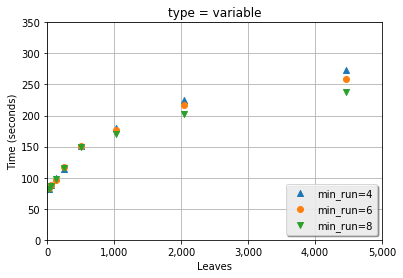}
	\includegraphics[width=225pt]{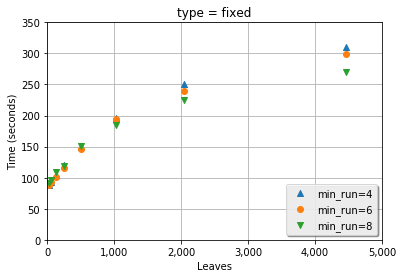}
	\caption{Effect of varying number of leaf nodes on time taken: \texttt{t5} vs. \texttt{t5}}
	\label{fig:t5_t5_time}
\end{figure*}

To construct the tree, the smallest number of leaf nodes was 32. Following this, the number of leaf nodes was doubled each time (maintaining a balanced tree). The exception was that after the experiment with 2,048 leaf nodes, 4,457 leaf nodes were used for the final run, thereby representing a single file from the corpus in each leaf.

The aims of this experiment were:
\begin{enumerate}
   \item Evaluate the effectiveness of the HBFT approach for exact matching (i.e. finding identical files) using recall.
   \item Identify an appropriate value for \texttt{MRSH-v2}'s \texttt{min\_run} parameter.
   \item Investigate the relationship between the size of the tree and the time taken to build and search the tree.
   \item Investigate the relationship between the size of the tree and the number of pairwise comparisons that are required to calculate a similarity score.
\end{enumerate}

When running the experiment, it became apparent that the first two aims are linked. Table~\ref{tab:recall} shows the recall associated with three values of \texttt{min\_run}: 4, 6 and 8. Using a \texttt{min\_run} value of 4 resulted in full recall. However, increasing \texttt{min\_run} to 6 or 8 resulted in a small number of files being omitted. When \texttt{min\_run} is set to 8, three files are not found in the tree. This indicates the dangers inherent in requiring longer matching runs. The files in question are \texttt{000462.text}, \texttt{001774.html}, \texttt{003225.html}. These files are 6.5 KiB, 6.6 KiB and 4.5 KiB in size respectively. Although each chunk is approximately 160 bytes, this is variable depending on the file content. While these are relatively small files, they are not the smallest in the corpus. This shows that even when the file is large enough to contain 8 chunks of the average size, a \texttt{min\_run} requirement of 8 successive matches may still not be possible. Similarly, using 6 as the \texttt{min\_run} value results in two files being missed. The type of HBFT used did not alter these results.

\begin{table}[h]
	\caption{Effect of min\_run on recall: identical files.} \label{tab:recall}
	\centering
	\begin{tabular}{R{1.1cm}R{1.8cm}}
		\tline
		min\_run & Recall \\
		\mline
		4 & 100\% \\
		6 & 99.96\% \\
		8 & 99.93\% \\
		\bline
	\end{tabular}
\end{table}

It should be acknowledged that if the aim is solely to identify identical files, then existing hash-based techniques will take less time and yield more reliable results. Intuitively, however, a system that is intended to find similar files should also find identical files. While the chunk size of 160 bytes will always fail to match very small files, it is desirable to find matches when file sizes are larger.

Figure~\ref{fig:t5_t5_time} shows the time taken to build the tree and search for all files. As the number of leaf nodes in the tree increases, so too does the time taken to search the tree. Higher values of \texttt{min\_run} use slightly less time, due to the fact that it is more difficult for a search to descend to a lower level when more matches are required to do so. However, as the recall for these higher values is lower, 4 was used as the \texttt{min\_run} value for further experiments.

\begin{figure*}[b]
	\centering
	\includegraphics[width=210pt]{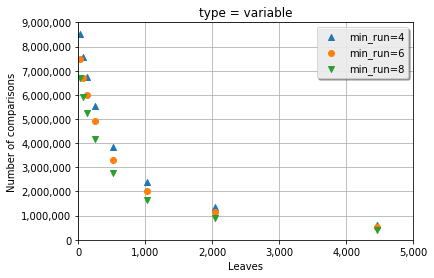}
	\includegraphics[width=210pt]{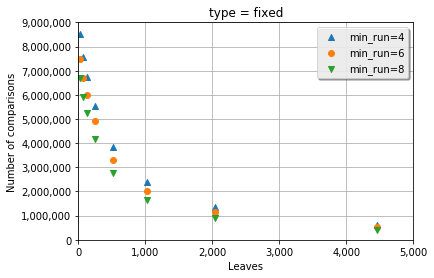}
	\caption{Effect of varying number of leaf nodes on number of comparisons: \texttt{t5} vs. \texttt{t5}}
	\label{fig:t5_t5_comparisons}
\end{figure*}

The times shown here relate only to building the tree and searching for files within it, and do not include the time for the pairwise comparisons at the leaves. Therefore, although using 32 leaf nodes results in the shortest search time (due to the shallower tree), it would require a most comparisons, as each leaf node represents $\frac{1}{32}$ of the entire corpus. As an illustration, using a variable-width tree with 32 leaf nodes and \texttt{min\_run} value of 4 requires 8,538,193 pairwise comparisons after searching the tree. A similar tree with 4,457 leaves requires 617,860 comparisons.

One issue that is important to note is that the time required to perform a full pairwise comparison is 319 seconds. However, for the largest trees, the times for building and searching the tree are 274 and 309 seconds for a variable and fixed  HBFT respectively. Thus, for a relatively small collection such as this, the use of the tree is unlikely to provide benefits in terms of overall running time.

Figure~\ref{fig:t5_t5_comparisons} plots the number of leaf nodes against the total number of comparisons required to complete the investigation. As the size of corpora increases, so does the number of pairwise comparisons required by \texttt{MRSH-v2}. Thus reducing this search space is the primary function of the tree. Larger trees tend to result in a smaller number of comparisons. For the largest trees (with 4,457 leaves), the \texttt{min\_run} value does not have a material effect on the number of comparisons required, regardless of whether the tree is variable-width or fixed-width. This implies that although searches tend to reach deeper into the tree (hence the longer running time), they do not reach substantially more leaves.

From this experiment, it can be concluded that using a \texttt{min\_run} value of 4 is desirable in order to find exact matches. This causes the time taken to search to be slightly longer, while having a negligible impact on the number of pairwise comparisons required afterwards. Fixed-width and variable-width HBFTs exhibit similar characteristics for a corpus of this size

\subsubsection{Experiment 2: \texttt{t5} vs. \texttt{win7} and \texttt{win7} vs. \texttt{t5}} \label{sec:experiments2}

The second experiment was designed to operate with larger dataset sizes. In this experiment, \texttt{t5} was used as a proxy for a set of known-illegal files, and \texttt{win7} was used to represent a seized disk.

The aims of this experiment were:
\begin{enumerate}
   \item Investigate whether the HBFT should represent the smaller or larger corpus.
   \item Contrast the performance of variable-width and fixed-width trees.
   \item Measure the effect on overall running time of using a HBFT.
\end{enumerate}

\begin{figure*}[t]
	\centering
    \includegraphics[width=210pt]{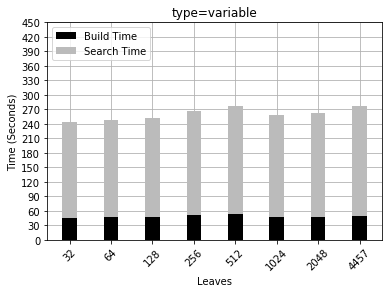}
    \includegraphics[width=210pt]{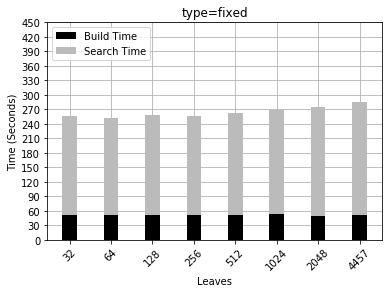}
    \caption{Time to search for \texttt{win7} in a \texttt{t5} tree.} \label{fig:t5_win7_time}
    \includegraphics[width=210pt]{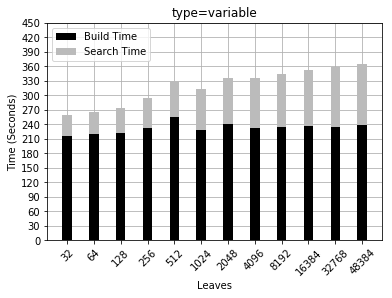}
    \includegraphics[width=210pt]{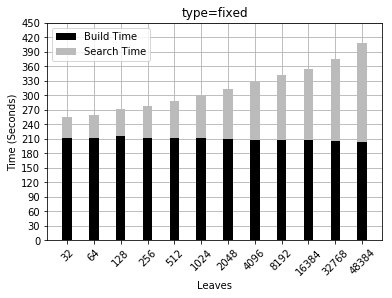}
	\caption{Time to search for \texttt{t5} in a \texttt{win7} tree.} \label{fig:win7_t5_time}
\end{figure*}

In pursuit of the first objective, the experiment was first run by building a tree to represent \texttt{t5} and then searching for the files contained in \texttt{win7}. The number of leaf nodes in this tree was varied in the same way as in Experiment~1. Then this was repeated by inserting \texttt{win7} into a tree and searching for the files from \texttt{t5}. Again the number of leaf nodes was doubled every time, with the exception that the largest tree contained one leaf node for every file in the collection (i.e. 48,384 leaves). This procedure was followed for both forms of tree.

The time taken to build and search the trees are shown in Figures~\ref{fig:t5_win7_time} and~\ref{fig:win7_t5_time}. Figure~\ref{fig:t5_win7_time} shows the results when the tree represents \texttt{t5}, with the time subdivided into the time spent building the tree and the time spent searching for all the files from \texttt{win7}. The total time is relatively consistent for both types of tree. This is unsurprising in the context of disjoint corpora. Most files will not match, so many searches will end at the root node, or at an otherwise shallow depth.

Figure~\ref{fig:win7_t5_time} shows results when the tree models \texttt{win7}. With only 32 leaf nodes, it is notable that all four experimental runs take approximately the same total time, regardless of the type of tree or the dataset that is chosen for the tree to represent. Due to its size, the build times for the \texttt{win7} trees are substantially longer than for \texttt{t5}. The search time exhibits a generally upward trend as the number of leaf nodes increases: a trend that is far more pronounced for the fixed-width tree.

This is a consequence of the hardware constraints associated with the setup of the experiment. Because memory footprint is constrained, a tree with 48,384 leaf nodes will contain Bloom filters that are much smaller than for trees with fewer nodes. For the variable-width tree representing \texttt{win7}, although its leaf nodes are 8KiB in size, its root node is 512MiB. In the corresponding fixed-width tree, the Bloom filters are all 64KiB. The false positive rate associated with Bloom filters is much higher for smaller Bloom filters. Thus even where two corpora have no files in common, searches int he fixed-width tree will descend deeper due to false positives higher in the tree, hence increasing the search time. This is likely to be even more pronounced in corpora that have a substantial number of similar files. Therefore, the fixed-width tree in its current design is unlikely to successfully scale to very large corpora.

Overall, the total time taken is less when the tree represents the smaller dataset. As with the first experiment, the total number of pairwise comparisons decreases as the number of leaves increases. Table~\ref{tab:comparisons} shows the total number of comparisons that are required when using the largest number of leaf nodes (i.e. 4,457 when the tree represents \texttt{t5} and 48,384 when \texttt{win7} is stored in the tree). Both types of tree require a smaller number of comparisons when the tree models \texttt{t5}. This, combined with the lower build and search time suggest that the preferred approach should be to use the smaller corpus to construct the HBFT.

\begin{table}[htb]
	\centering
	\caption{Number of pairwise comparisons required for largest trees: \texttt{t5} vs. \texttt{win7}} \label{tab:comparisons} 
   	\begin{tabular}{L{1.2cm}L{1cm}R{1.6cm}R{1.8cm}}
    \tline
    Tree & Search & Fixed & Variable \\
    \mline
    \texttt{t5} & \texttt{win7} & 98,260 & 98,260 \\
    \texttt{win7} & \texttt{t5} & 193,924 & 101,386 \\
    \bline
    \end{tabular}
\end{table}

Memory is an additional consideration. Using a HBFT to model the larger dataset requires the similarity hashes of all its files to be cached at the leaves. This requires more memory than for the smaller collection, thus reducing the amount of memory available to store the HBFT itself. 

Following these observations, the experiment was repeated once more. A variable-width tree was used, which modelled \texttt{t5} with 4,457 leaves. All files from \texttt{win7} were then searched for. The total running time, including pairwise comparisons, was 1,094 seconds. In comparison, the time taken to perform a full pairwise comparison using \texttt{MRSH-v2} is 2,858 seconds. 

\subsubsection{Experiment 3: Planted evidence} \label{sec:experiment3}

The final experiment involved overlapping datasets, constructed as follows:

\begin{itemize}
   \item A set of simulated ``known-illegal'' files: 4,000 files from \texttt{t5}.
   \item A simulated seized hard disk: the \texttt{win7} image, plus 140 files from \texttt{t5}, as follows:
   \begin{itemize}
      \item 100 files that are contained within the 4,000 ``illegal'' files.
      \item 40 files that themselves are not contained within the ``illegal'' files, but that have a high similarity with files in the corpus, according to \texttt{MRSH-v2}. 10 of these files have a similarity of 80\% or higher, 10 have a similarity between 60\% and 79\% (inclusive), 10 have a similarity between 40\% and 59\% (inclusive) and 10 have a similarity between 20\% and 39\% (inclusive).
   \end{itemize}
\end{itemize}

The aims of this third experiment were:
\begin{enumerate}
   \item Evaluate the time taken to perform a full search, compared with the all-against-all pairwise approach of \texttt{MRSH-v2}.
   \item Evaluate the success of the approach in finding the 100 ``illegal'' files that are included verbatim in the hard disk image, and the 40 files from the image that are similar to ``illegal'' files, according to \texttt{MRSH-v2}.
\end{enumerate}

For the first aim, the primary metric is the time taken for the entire process to run, comprising the time to build the tree, the time to search the tree and the time required to conduct the pairwise comparisons at the leaves. In evaluating the latter aim, recall is used. Here, ``recall'' refers to the percentage of the 100 identical files that are successfully identified, and ``similar recall'' refers to the percentage of the 40 similar files that are successfully found. A file is considered to have been found if the search for the file it is similar or identical to reaches the leaf node that contains it, yielding a pairwise comparison.

\begin{figure*}[htb]
	\centering
    \includegraphics[width=210pt]{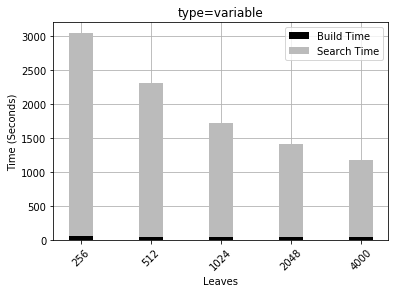}
    \includegraphics[width=210pt]{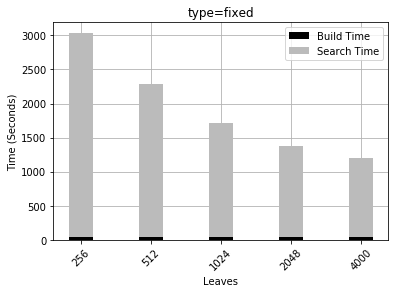}
    \caption{Time to search for planted evidence (including pairwise comparisons).} \label{fig:t5_win7_plant_time}
\end{figure*}

The total running time for \texttt{MRSH-v2} was 2,592 seconds. The running times of the HBFT approach are shown in Figure~\ref{fig:t5_win7_plant_time}. Following the insights gained in the previous experiment, the smaller collection of 4,000 ``illegal'' files was used to construct the tree and then searches were conducted for all of the files in the larger corpus. The  ``Search Time'' includes the time spent searching the tree and the time to perform the comparisons at the leaves.

As expected, for both types of tree the maximum number of leaf nodes resulted in the fastest run time. This configuration also yielded the maximum reduction in the number of pairwise comparisons required, without substantially adding to the time required to build and search the tree. The remainder of this analysis focuses on this scenario, where the tree has 4,000 leaf nodes.

Using a variable-width tree took 1,182 seconds (a 54\% reduction in the time required for an all-against-all pairwise comparison). The fixed-width tree took 1,207 seconds (a 53\% reduction). This illustrates that in terms of run-time, the HBFT approach offers substantial speed gains over pairwise comparison. Due to the lack of scalability of the pairwise approach, this difference is likely to be even more pronounced for larger datasets.

\begin{table}[htb]
	\centering
	\caption{Similar recall for Planted Evidence experiment.} \label{tab:planted_approx_recall} 
   	\begin{tabular}{L{1.8cm}C{1.1cm}C{1.1cm}C{1.7cm}}
    \tline
    \texttt{MRSH-v2} & Files & Files & Similar \\
    similarity & planted & found & recall \\
    \mline
    80\%-100\% & 10 & 10 & 100\% \\
    60\%-79\% & 10 & 10 & 100\% \\
    40\%-59\% & 10 & 10 & 100\% \\
    20\%-39\% & 10 & 8 & 80\% \\
    \mline
    \textbf{Overall} & 40 & 38 & 95\% \\
    \bline
    \end{tabular}
\end{table}

In terms of effectiveness, all 100 files that were common to the two corpora were successfully found in both tree types. The similar recall is shown in Table~\ref{tab:planted_approx_recall} and is the same for both types of tree. All files with a \texttt{MRSH-v2} similarity of 40\% or greater with a file in the ``illegal'' set were successfully identified. Two files with a lower similarity (25\% and 26\%) were not found. This yields an overall similar recall score of 95\% for all 40 files. 

This is an encouraging result, indicating that the HBFT approach is extremely effective at finding files that are similar above a reasonable threshold of 40\% and exhibits full recall for identical files. Thus it can be concluded that the HBFT data structure is a viable alternative to all-against-all comparisons in terms of effectiveness, while achieving substantial speed gains.

\section{Conclusions and Future Work} \label{sec:conclusions}

This paper aimed to investigate the effectiveness of using a Hierarchical Bloom Filter Tree (HBFT) data structure to improve upon the all-against-all pairwise comparison approach used by \texttt{MRSH-v2}. A number of experiments were conducted with the aim of improving the speed of the process. Additionally, it was important that files that should be found were not omitted (i.e. that recall is maintained).

The first experiment found that while HBFTs with more leaf nodes take longer to build and search, they reduce the number of pairwise comparisons required by the greatest degree. It also suggested the use of a \texttt{min\_run} value of 4, as higher values resulted in imperfect recall for identical files.

The results of the second experiment indicated that when using corpora of different sizes, it is preferable to build the tree to model the smaller collection and then search for the files that are contained the larger corpus. For larger trees, it was additionally noted that the fixed-width HBFT did not scale as well as its variable-width counterpart. This is due to the small size of the Bloom filters used in the tree as the number of nodes increases.

For the final experiment, a Windows 7 image was augmented by the addition of a number of files that were identical to those being searched for, and a further group that were similar. The HBFT approach yielded a recall level of 100\% for the identical files and of 95\% for the similar files, when using \texttt{mrsh-v2} as ground truth. On examining the two files that were not found, it was noted that these had a relatively low similarity to the search files (25\% and 26\% respectively), with all files with a higher similarity score being identified successfully. The run time for this experiment was substantially quicker than an all-against-all comparison: a 54\% time reduction for a variable-width tree and a 53\% reduction for a fixed-width tree.

These experiments lead to the conclusion that the HBFT approach is a highly promising technique. Due the poor scalability of the traditional all-against-all approach, it can be inferred that this performance improvement will be even more pronounced as datasets become larger.

Given the promising results of the experiments presented in this paper, further work is planned. A number of avenues for future work are apparent:
\begin{itemize}
   \item Cuckoo filters have been identified as a promising replacement for Bloom filters for approximate matching purposes~\cite{Gupta2015}. It is possible that these could be incorporate into a similar hierarchical tree structure to produce further improvements.
   \item Currently, when building the tree, files are allocated to leaf nodes in a round-robin fashion. For trees with multiple files represented at each leaf, it may be possible that a more optimised allocation mechanism could be used for this (e.g. to allocate similar files to the same leaf node).
   \item The current model also uses balanced trees, with the result that all successful searches reach the same depth in the tree. Further investigation may reveal circumstances where an unbalanced tree is preferable so as to shorten some more common searches.
   \item Parallelisation and distribution are highly likely to yield further performance improvements, and this should be investigated. 
   \item Fixed-width HBFTs do not scale to the same extent as variable trees, due to the high false positive rates that are associated with the small Bloom filters that result from using large trees with many nodes. Although the experiments presented in this paper indicate that variable-width HBFTs are preferable, there may be circumstances where fixed-width trees may be useful, due to their theoretical advantages noted in Section~\ref{sec:hbft}.
   \item While these experiments have used \texttt{MRSH-v2} as the algorithm for calculating the similarities at the leaf nodes, other algorithms should be considered also (e.g. \texttt{sdhash}).
\end{itemize}

\bibliographystyle{apacite}
\balance
\bibliography{bibfile}

\end{document}